# Locally resolved electronic textures of reconstruction domains in marginally twisted monolayer–bilayer graphene

Sean M. Walker[1+], Patrick Sarsfield[2,3+], Isaac Soltero[2,3+], Xue-Ying LiYang[1+], Laurent Molino[1], Ryan Plumadore[1], Kenji Watanabe[4], Takashi Taniguchi[5], Vladimir Fal'ko[2,3]*, Adina Luican-Mayer[1,6]*

[1] *Department of Physics, University of Ottawa, Ottawa, Ontario, K1N 9A7, Canada,*
[2] *Department of Physics and Astronomy, University of Manchester, Manchester, UK*
[3] *National Graphene Institute, University of Manchester, Manchester, UK*
[4] *Research Center for Electronic and Optical Materials, National Institute for Materials Science, 1-1 Namiki, Tsukuba 305-0044, Japan*
[5] *Research Center for Materials Nanoarchitectonics, National Institute for Materials Science, 1-1 Namiki, Tsukuba 305-0044, Japan*
[6] *University of Illinois, Chicago, USA*
*luican-mayer@uottawa.ca *aluica2@uic.edu *vladimir.falko@manchester.ac.uk

[+] *These authors contributed equally to the work*

Controlling the stacking and rotational registry of graphene layers provides a powerful handle on atomic-scale structural reconstructions that alter the electronic landscape at the nanoscale. In particular, this governs how massless and massive Dirac fermions coexist and interact at the monolayer-bilayer graphene interface. In the limit of marginal twist, the system reconstructs into domains of distinct vertical stacking order, introducing characteristic electronic properties and new electronic length scales, a regime that, despite its structural richness, remains largely unexplored. Here, using scanning tunnelling microscopy and spectroscopy, we demonstrate that at very low rotation angles the monolayer–bilayer graphene system relaxes into a network of three distinct stacking domains with individual electronic textures revealed through spatially resolved spectroscopic mapping and corroborated by computed local density of states. We report switching of the hierarchy of the tunnelling characteristics between Bernal and rhombohedral domains as a function of bias voltage. Furthermore, the measured spectroscopic maps exhibit theoretically anticipated domain wall 'twirling' around energetically unfavorable AAB stacking nodes, promoted by out-of-plane deformations. Our results shed light on fundamental structure–property relationships underpinning moiré-driven phenomena, opening new avenues for harnessing structural degrees of freedom in van der Waals heterostructures.

Vertically stacked systems of 2D materials offer a promising platform for band structure engineering due to their sensitivity to twist angle and the wide range of possible structures available in multilayer systems. While much attention has focused on the magic-angle regime — where dispersionless flat bands in twisted bilayer graphene graphene[1] and correlated states in more complex multilayer systems emerge from moiré band reconstruction[2,3] - the physics of the marginally twisted regime is qualitatively distinct [4,5] and comparatively underexplored.

At twist angles very near 0°, the system is no longer governed by a uniform moiré potential. Instead, long moiré periods drive lateral reconstruction into domains of uniform stacking separated by domain walls, a structural transformation observed by scanning probe and transmission electron microscopy in twisted bilayer graphene [5-10] and in more complex twisted multilayers [11-16]. In twisted monolayer-bilayer graphene at somewhat larger angles (0.9°–1.6°), transport experiments have revealed electric-field tunable correlated states in topologically nontrivial flat bands near the Dirac point [17-21]. In the marginal twist regime, however, the stabilization of otherwise unstable stacking domains introduces a fundamentally different platform: one where electronic structure is governed by domain size, reconstruction-induced textures, and emergent length scales rather than by a moiré potential. The electronic properties of these reconstruction domains and their influence on local electronic structure remain largely unexplored.

In this work, we realize marginally twisted monolayer graphene on bilayer graphene, leading to structural reconstruction into a network of distinct stacking domains. Using scanning tunnelling microscopy and spectroscopy (STM/STS), we directly image these domains and resolve their electronic textures with nanometre-scale spatial resolution.

We perform STM/STS on marginally twisted monolayer (1L)/ bilayer (2L) heterostructures placed on a hBN substrate, as schematically represented in Figure 1 (a), where the twist angle θ is very close to 0°. Figure 1 (b) shows a typical scanning tunneling microscopy topograph acquired on the sample. The bright dots correspond to the moiré pattern that emerges between the monolayer and the bilayer at such small angles. From the period of this superlattice ($\lambda \sim 47 \pm 2$ nm) we extract a relative twist angle $\theta = 0.30° \pm 0.01°$. The deviation observed in parts of the sample is a consequence of rotation angle inhomogeneity, which can be attributed to strain[22,23] as detailed in Supplementary Figure 1.

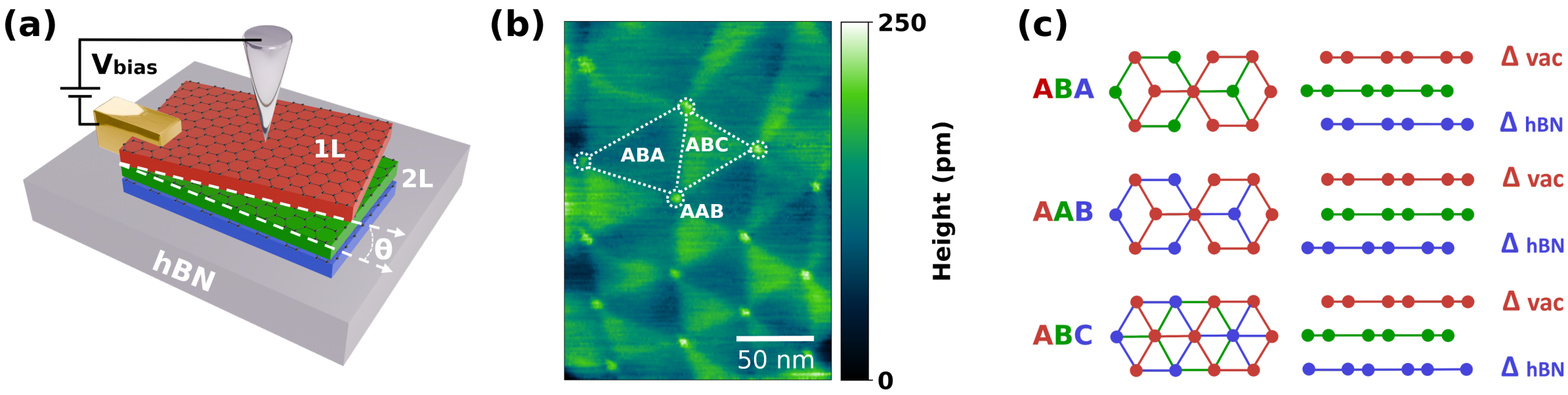

**Figure 1**: (a) Schematic of the STM measurement of a monolayer (1L)/bilayer graphene (2L) vertical heterostructure on the surface of hexagonal boron nitride (hBN). The twist angle between 1L and 2L is θ = 0.3°. (b) STM topograph showing triangular reconstruction domains ($V_B$ = 250mV; $I_t$ = 350pA). The different stackings identified are indicated within the dashed lines. (c) Schematic of the local stacking orders associated with the triangular domains (ABA and ABC) and the domain network nodes (AAB). The associated on-layer proximity induced potentials are labelled as ΔhBN for graphene-hBN interface and Δvac for graphene-vacuum interface.

The triangular domains originate in the different high symmetry stacking regions within the moiré pattern, which expanded to reach an energetically favorable configuration. Figure 1 (c) details the vertical arrangement for the three structures ABC, AAB, and ABA. We assign the regions in the topographic maps to these three stackings as indicated in Figure 1 (b). The nodes of the domain wall network are associated with the less favorable AAB stacking, while the triangles are associated with ABA and ABC. Consistent with previous reports, the STM topographic maps of rhombohedral stacking (ABC) regions have a greater apparent height compared to Bernal stacking regions (ABA) as shown in Supplementary Figure 2. This assignment is further validated by examining the density of states, as discussed below.

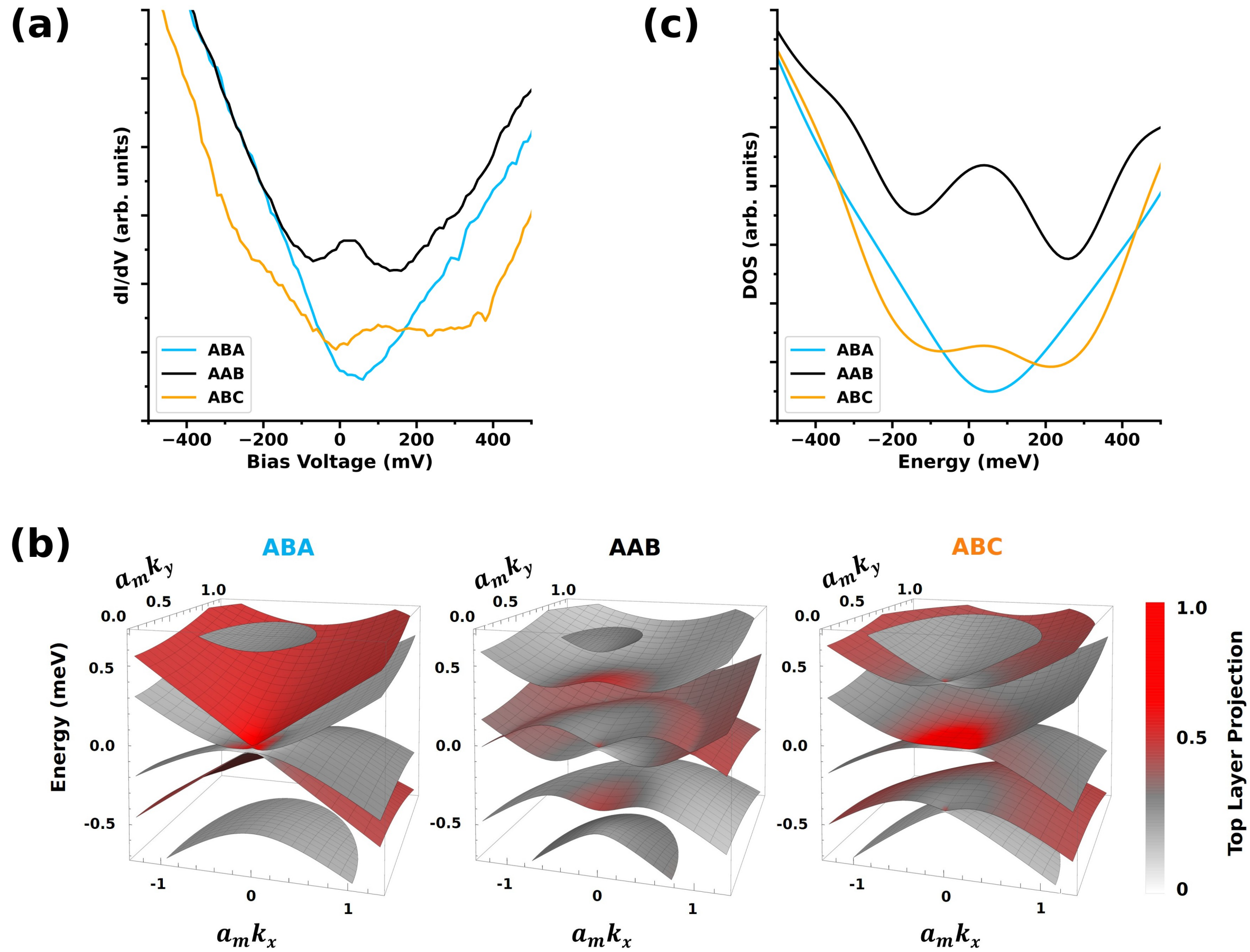

**Figure 2:** (a) Scanning tunneling spectra (STS) acquired on the regions of three high symmetry stackings, as indicated. (b) Band dispersion calculated from the hybrid $k{\cdot}p$ tight-binding model for the three high-symmetry stacking configurations with 82meV thermal broadening and on-layer proximity induced potentials ΔhBN and Δvac. The color scale indicates the projection of each state onto the topmost graphene layer. (c) Calculated density of states of three high symmetry stackings, as indicated.

Figure 2(a) presents scanning tunnelling spectroscopy measurements acquired on each of the three high-symmetry stacking configurations. Using a hybrid k.p tight-binding model[24] for trilayer graphene we computed the density of states projected onto the topmost graphene layer, which is proportional to the tunnelling current measured in STS (see Supplementary Note 3). We implemented self-consistent Hartree corrections to account for electrostatic interactions between carriers. The corresponding electronic band dispersions are shown in Figure 2(b). We included two on-layer proximity induced potentials in our Hamiltonian (see Supplemental Note 3), as illustrated in Figure 1(c). The first, $\Delta hBN = 18meV$[25], accounts for the effect of the graphene–hBN interface, while the second, $\Delta vac$, represents the graphene–vacuum interface at the top surface and was treated as a fitting parameter. We set $\Delta vac = 75meV$ to achieve the best agreement between the calculated density of states and the STS spectra shown in Figure 2(a). All other parameters in our Hamiltonian follow the Sloncewski-Weiss-McClure parametrization.[24,26,27]

For a direct comparison with the STM spectra we compute the local density of states taking into account thermal broadening of the Fermi function in graphene and the tip by 82 meV (see Supplementary Note 4), which smears out various fine-energy features in the electron spectra (such as mini gaps at the respective subband edge) and hides the effect of the pseudomagnetic field and scalar potential energy shift induced by lattice reconstruction effects (see Methods and Supplementary Note 5). For example, for ABA stacking both modelling and room-temperature STM manifest typical V-shaped DoS profile (linked to the Dirac type band painted red in Fig. 2(b)) without any fine features specific for conduction/valance band edges, in contrast to what has been observed in low-temperature STM studies[28] .

AAB stacking regions also display V-shape spectra, but with a characteristic peak at the Fermi level, in agreement with the theoretical calculations. For the ABC stacking, experimentally, the most prominent feature is a conduction band peak, and it is consistent with the density of states calculated for ABC stacking and corresponds to a pair of flat surface bands[15,21,29-31]. However, the energy of the experimentally observed feature on the conduction band edge is ∼50meV deeper in the band than predicted. The corresponding density of states are plotted in Supplementary Note 6.

Figure 3 illustrates the spatial variation of the domain structure and its associated electronic properties across adjacent stacking domains. In Figure 3(a), the experimental STM topographic map, displayed as a 3D plot, highlights the large apparent height at the moiré sites (AAB stacking). To track how spectral features change across domains, we performed STS measurements along a line crossing the domain boundaries. The resulting spectroscopic line map is shown in Figure 3(b), crosses from the ABC stacking triangle, through the domain wall (DW), and into the ABA triangle, as indicated in the color plot. The color scale represents measured dI/dV, proportional to the local density of states, where blue contrast denotes higher intensity. We observe that the features seen in the point spectra of Figure 2(a) remain uniform within each domain, within our experimental resolution, and change systematically as the stacking sequence transitions between domains.

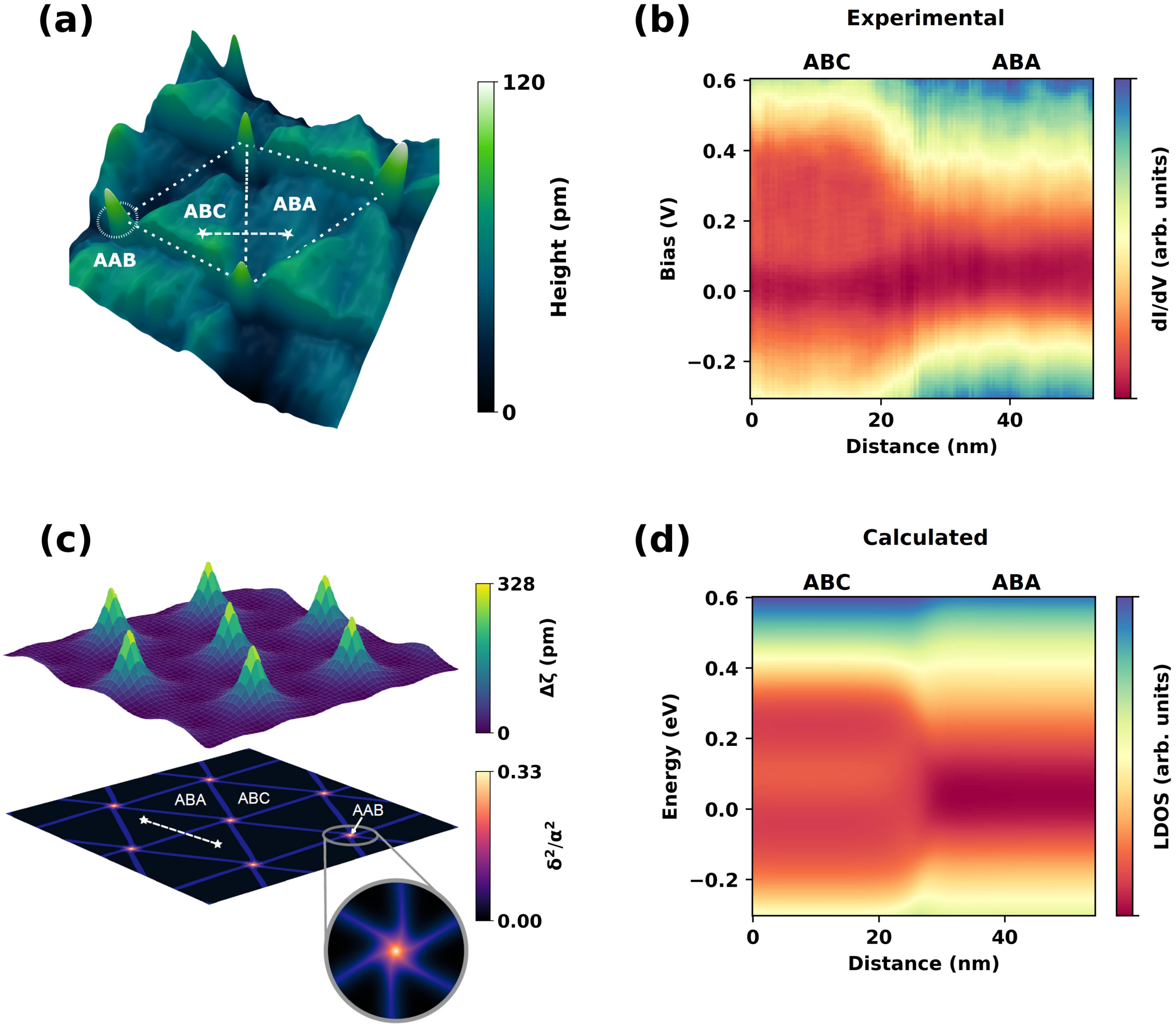

**Figure 3:** (a) STM topographic map showing apparent heights of three stacking regions. The dashed line crossing the domain boundaries represents the trajectory of the spectroscopic line map in (b). (b) Spectroscopic line map across a domain wall. (c) Theoretical lattice reconstruction showing the out-of-plane displacement and local lattice relaxation within the top layer. The top panel shows the top layer displacement (similar for the other two layers), while the bottom panel presents the displacement parameter δ, with the inset highlighting the "twirling" at dislocation network nodes. (d) Theoretically calculated STS line map along the trajectory indicated by the dashed line in (c) across a domain wall.

To understand this behavior, we performed mesoscale lattice relaxation calculations accounting for in-plane and out-of-plane deformations of each layer over the moiré unit cell, resulting in the formation of ABC and ABA stacking domains (see Supplementary Note 7). Interestingly, we find that the minimal energy configuration of the system is when the three layers are bent in the same direction at the AAB stacking nodes, with a relative height of ~330 pm with respect to the ABC and ABA domains as shown in top panel of Figure 3(c). Moreover, this corrugated state is accompanied by 'twirling' of the domain wall network around the AAB nodes[32-35]. To visualize this effect, we estimate the deviation parameter $\delta$ (bottom panel in Figure 3(c)), defined as the minimum of the lateral offset between atoms of opposite sublattices between the lower and upper layers at the twisted interface, such that $\delta = 0$ for perfect ABC and ABA stackings and $\delta = a/\sqrt{3}$ for AAB stacking, where $a$ is the monolayer lattice parameter. To facilitate comparison with experiment, we compute the spectroscopic line map along an equivalent trajectory connecting stacking domain centers across a domain wall, following the dashed line indicated in Figure 3(c). The resulting theoretical map, shown in Figure 3(d), reproduces the key features of the experimentally observed electronic structure. In particular, we extract the domain wall width in each case as the distance between the two regions with constant dI/dV (or LDOS). This yields a domain wall width of $(8 \pm 3)$ nm for the experimental data (see Supplementary Note 8), consistent with previous reports of stacking boundary widths[36,37] and in good agreement with the width of approximately 7 nm obtained from our theoretical calculations.

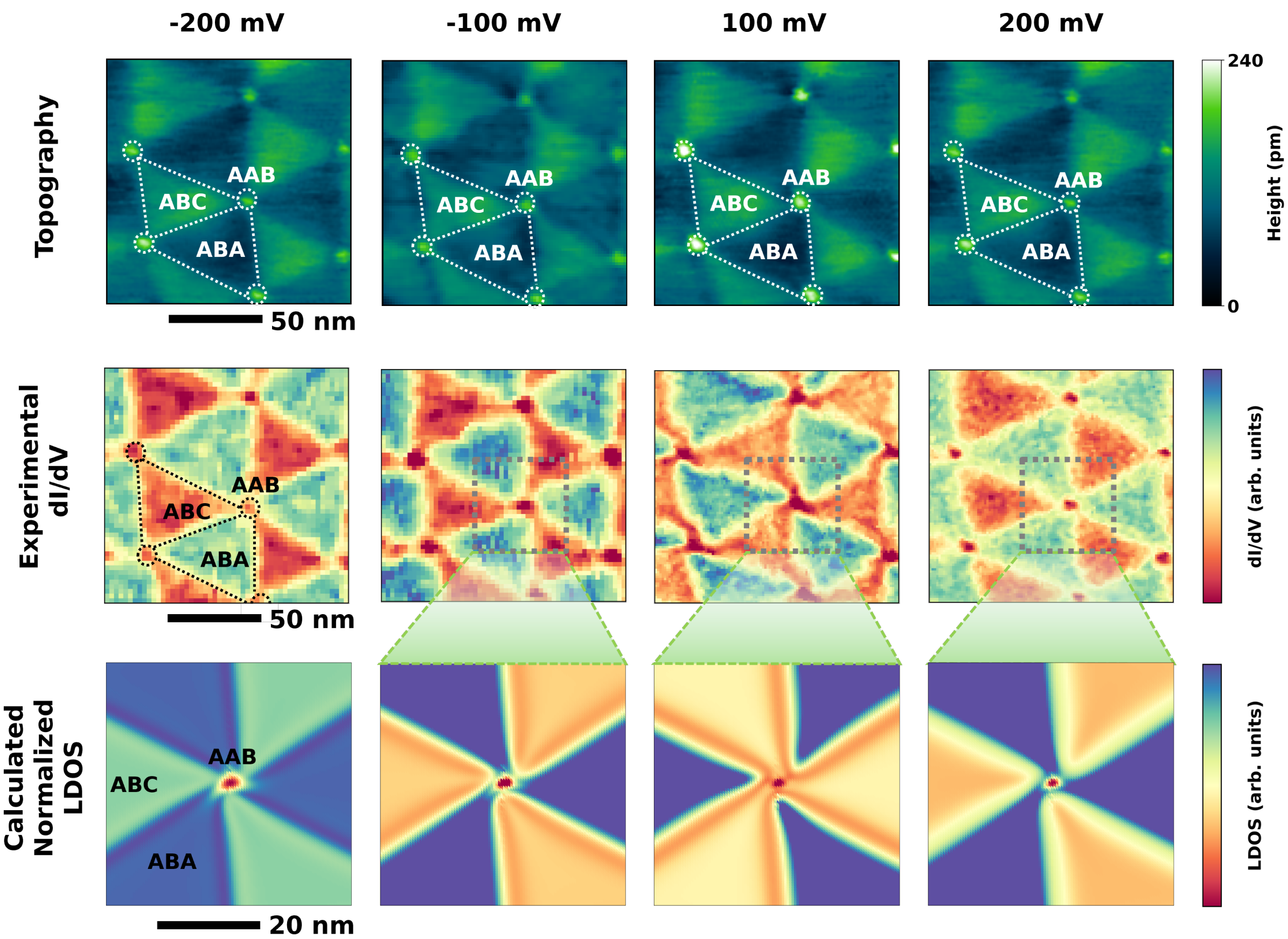

**Figure 4:** Top row: Sequence of STM topographic maps acquired at the indicated bias voltages (It = 150 pA). Middle row: Experimentally measured spectroscopic dI/dV maps corresponding to the labelled bias voltages. Bottom row: Theoretically calculated LDOS maps, normalized to reflect constant-current measurement conditions. The stacking domains are indicated as a guide to the eye.

We now explore the spatial distribution of the local density of states across a region with several different stacking domains. In the top row of Figure 4 we present topographic maps of the same region, acquired at different bias voltages, as indicated.

In interpreting the color contrast of these maps, we note that our STM operates in constant current mode, and the apparent height measured in STM topography reflects a convolution of surface topography and the local density of states integrated up to the applied bias. As a result, bias-dependent contrast in the topography can arise from variations in the electronic structure rather than purely structural height differences. In the middle row of Figure 4, we present the experimentally acquired dI/dV maps, a quantity proportional to the local density of states at the corresponding bias voltages. We observe inversion of domain contrasts: as the energy is varied from −100 mV to 100 mV, ABC domains which exhibit lower density of states (red) become higher in intensity (blue) and then revert to lower intensity again at 200 mV, with the opposite trend observed for ABA domains. This effect is captured in the STS shown in Figure 2(a), where the dI/dV intensity of the ABC domains is higher than that of the ABA domains in the low-bias part of the conduction band, and lower at other energies, consistent with the observed contrast inversion in the dI/dV maps.

However, we note disagreement at the AAB nodes, which show lower intensity compared to the surrounding triangular domains. This can be understood by noting that $dI/dV$ maps are acquired in constant-current mode, thus, the measured signal does not directly correspond to the LDOS given by:

$$\rho(E) = \int_{BZ} \sum_{n=1}^{6} \frac{d^2\boldsymbol{k}}{(2\pi)^2} |u_n^{A3}(\boldsymbol{k}) + u_n^{B3}(\boldsymbol{k})|^2 \delta(E - E_n(\boldsymbol{k})),$$

where $E$ is energy, $\boldsymbol{k}$ is wave vector, $n$ is band index and $u_n^{A3}(\boldsymbol{k}), u_n^{B3}(\boldsymbol{k})$ are components of the normalized wavefunction on sublattices $A$ and $B$ of the topmost graphene layer (layer 3). To account for this effect in our theoretical calculation, we normalize the LDOS by the total number of states:

$$N(\varepsilon) = \int_0^{\varepsilon} \rho(E) dE,$$

and calculate $\rho(\varepsilon)/N(\varepsilon)$ where $\varepsilon$ is related to bias voltage by $\varepsilon = eV_{bias}$. This quantity is proportional to experimentally measured $dI/dV$ across the sample measured with a constant current. The resulting normalized LDOS maps around AAB nodes are shown in the bottom panel of Figure 4. They reproduce the alternating relative intensity of the two triangular domains and show good agreement near the nodes. The maps also display similar spatial features: at a bias of 100 mV, the domain walls appear to connect around the nodes, whereas at other biases they remain separated. Furthermore, the domain wall “twirling” around the nodes, predicted by our lattice reconstruction calculations in Figure 3(b), is captured in both the $dI/dV$ maps and the normalized LDOS. This indicates that the twirling arises from lattice relaxation and the resulting out-of-plane deformations of the graphene layers.

In conclusion, we have examined the near-zero twist angle limit of monolayer–bilayer graphene, a regime dominated by atomic relaxation domains rather than periodic moiré potential. Through scanning tunneling microscopy and spectroscopy, supported by theoretical modeling, we show that the system settles into three structurally and electronically distinct stacking domains (ABA, ABC, AAB). We also report on the continuous evolution of electronic character as the system transitions between distinct stacking configurations and estimate the characteristic width of the domain walls. Spatially resolved spectroscopic mapping reveals that each domain carries its own electronic fingerprint, shaped by the local vertical stacking configuration, and that the relative prominence of Bernal and rhombohedral tunneling characteristics inverts as a function of applied bias. Notably, the domain wall network exhibits a twirling texture around AAB sites, driven by out-of-plane lattice deformations and in full agreement with theoretical predictions. Broadly, our findings highlight structural relaxation as a primary determinant of electronic behavior in twisted van der Waals systems, and that the marginal twist regime offers a largely untapped space for engineering electronic properties through stacking order.

## Methods

### Fabrication

The twisted mono-bilayer graphene was fabricated using the tear and stack and inverted transfer technique with a PPC (Polypropylene carbonate)/ PDMS(Polydimethylsiloxane) stamp and mechanically exfoliated flakes. A hexagonal boron nitride (hBN) flake was first picked up to be used to tear-and-twist graphene layers at 45 °C, and the graphene layers were picked up consecutively. The PPC film with the stack was then peeled off from the stamp and transferred onto a clean, gold-coated Si/$SiO_2$ substrate. The PPC was cleaned by annealing in a forming gas mixture of $H_2$/Ar (5:95) at 230°C for 10h and in vacuum ($\sim 10^{-6}$ Torr) at 300°C for 1h. To provide the electrical contact necessary for STM measurements, the hBN surface was covered with a layer of Au/Ti (50nm/5nm), which was patterned by electron beam lithography and deposited by thermal evaporation. Lastly, the multilayer graphene area was further cleaned by AFM contact mode brooming to remove any polymer residues.

**STM**

STM/STS experiments were conducted by a commercial RHK PanScan Freedom system at 300 K and $10^{-10}$ mbar, with a Pt/Ir STM tip. STS were acquired with a voltage modulation of 5 mV and frequency of 1.325 kHz and were performed at several areas across the sample with different tunneling junction settings as shown in Supplementary Figure 7.

**Mesoscale lattice relaxation calculation**

The modeling of atomic relaxation in trilayer graphene with the topmost layer twisted by an angle $\theta$ is performed through a multiscale approach[38,39] based on an interpolated adhesion energy density, dependent on both the vertical stacking and interlayer distance[40], and continuum elasticity theory. Within this approach, we introduce the in-plane, $\boldsymbol{u}^{(\ell)}$, and out-of-plane displacement fields, $\zeta^{(\ell)}$, in each layer ($\ell = 1, 2, 3$), such that these minimize the total energy of the moiré unit cell (mUC),

$$\mathcal{H} = \int_{\text{mUC}} \mathrm{d}^2 r \left[ W\left(\boldsymbol{r}_0^{(1,2)}, d^{(1,2)}\right) + W\left(\boldsymbol{r}_0^{(2,3)}, d^{(2,3)}\right) + \sum_{\ell=1}^{3} \{U_\ell + B_\ell\} \right]. \tag{1}$$

where

$$U_\ell = \frac{\lambda}{2}\left(u_{ii}^{(\ell)}(\boldsymbol{r})\right)^2 + \mu u_{ij}^{(\ell)}(\boldsymbol{r}) u_{ji}^{(\ell)}(\boldsymbol{r}),$$
$$B_\ell = \frac{\kappa}{2}\left[\left(\nabla^2 \zeta^{(\ell)}(\boldsymbol{r})\right)^2 + 2(1-\sigma)\left\{\left(\partial_x \partial_y \zeta^{(\ell)}(\boldsymbol{r})\right)^2 - \left(\partial_x^2 \zeta^{(\ell)}(\boldsymbol{r})\right)\left(\partial_y^2 \zeta^{(\ell)}(\boldsymbol{r})\right)\right\}\right]. \tag{2}$$

In Eq. (1), $W$ is the adhesion energy density, accounted for both interfaces between graphene monolayers, and is determined by the lateral offset between A sublattices in neighboring layers, $\boldsymbol{r}_0$, and the interlayer distance $d$. Explicit functional form of $W(\boldsymbol{r}_0, d)$, parametrized from density functional theory (DFT) calculations[40]. The terms $U_\ell$ and $B_\ell$ are the elastic and bending energy densities, given in terms of the strain tensor $u_{ij}^{(\ell)}(\boldsymbol{r}) = \frac{1}{2}\left[\partial_i u_j^{(\ell)}(\boldsymbol{r}) + \partial_j u_i^{(\ell)}(\boldsymbol{r}) + \left(\partial_i \zeta^{(\ell)}(\boldsymbol{r})\right)\left(\partial_j \zeta^{(\ell)}(\boldsymbol{r})\right)\right]$, first Lamé coefficient $\lambda = 450\text{eV/nm}^2$, shear moduli $\mu = 920\text{eV/nm}^2$, bending rigidity $\kappa = 1.44\text{eV}$, and Poisson ratio $\sigma = 0.165$ [41,42]. The lateral offset vectors for the AB stacked ($\ell = 1$ and 2) and the twisted interface ($\ell = 2$ and 3) are

$$\begin{aligned} \boldsymbol{r}_0^{(1,2)}(\boldsymbol{r}) &= \boldsymbol{r}_0^{(\mathrm{AB})} + \boldsymbol{u}^{(2)}(\boldsymbol{r}) - \boldsymbol{u}^{(1)}(\boldsymbol{r}) + \frac{d^{(1,2)}(\boldsymbol{r})}{2}\nabla\left[\zeta^{(1)}(\boldsymbol{r}) + \zeta^{(2)}(\boldsymbol{r})\right], \\ \boldsymbol{r}_0^{(2,3)}(\boldsymbol{r}) &= \theta\hat{\boldsymbol{z}} \times \boldsymbol{r} + \boldsymbol{u}^{(3)}(\boldsymbol{r}) - \boldsymbol{u}^{(2)}(\boldsymbol{r}) + \frac{d^{(2,3)}(\boldsymbol{r})}{2}\nabla\left[\zeta^{(2)}(\boldsymbol{r}) + \zeta^{(3)}(\boldsymbol{r})\right], \end{aligned} \tag{3}$$

where $\boldsymbol{r}_0^{(\mathrm{AB})} = \frac{a}{\sqrt{3}}\hat{\boldsymbol{y}}$. Similarly, the interlayer distances are

$$d^{(k,l)}(\boldsymbol{r}) = d_0 - \zeta^{(k)}(\boldsymbol{r}) + \zeta^{(l)}(\boldsymbol{r}), \tag{4}$$

with $d_0 = 0.339\text{nm}$ an interlayer distance reference. The fourth term in Eqs. (3) is an additional in-plane offset shift due to bending of the layers, which plays an important role for non-encapsulated structures. Note that bending the layers in opposite directions cancels the contribution of this term, while bending in the same direction results in an effective lateral offset, contributing to atomic reconstruction effects. Furthermore, given the low bending rigidity of graphene monolayers, vertical displacement of the layers in the same direction results in a drastic reduction of the AA stacking node area, accompanied by twirling of domain walls around this region.

The optimal configuration of the displacement fields was found by defining these on a grid on the mUC, where the spacing between points was reduced until achieving convergence. The minimal energy was determined using the limited-memory Broyden-Fletcher-Goldfarb-Shanno algorithm (L-BFGS) with a tolerance of $10^{-11}$eV. The results for displacements fields in the trilayer system with $\theta = 0.3°$ are shown in Supplementary Figure 3.

## Acknowledgements

A.L.-M., S.M.W., X.-Y.L. and R.P. acknowledge support from the Natural Sciences and Engineering Research Council of Canada (NSERC) Discovery Grant (RGPIN-2022-05215) and the Alliance Quantum Consortium Grant (ALLRP/578466-2022). A.L.-M. acknowledges support from the Canadian Institute for Advanced Research (CIFAR). P.S. acknowledges support from the CDT Graphene-NOWNANO. V.F. acknowledges support from the British Council ISPF grant (1185409051). I.S. acknowledges financial support from the University of Manchester Dean's Doctoral Scholarship. K.W. and T.T. acknowledge support from the CREST (JPMJCR24A5), JST and World Premier International Research Center Initiative (WPI), MEXT, Japan.

**Author contributions**

A.L.-M. and V.F. conceived and supervised the project. and A.L.-M. S.M.W., X.-Y.L., R.P. and L.M. performed the experiments and acquired the data. S.M.W. analysed the data, with contributions from X.-Y.L. and L.M. I.S, P.S. and V.F. developed the theoretical model. K.W. and T.T. provided hBN crystals. All authors discussed the results and contributed to writing the manuscript.

**Supplementary information for: Locally resolved electronic textures of reconstruction domains in marginally twisted monolayer–bilayer graphene**

## Supplementary Note 1: Twist angle variations due to strain

As stated in the main text, the spatial variations of the local twist angle across the sample, shown in (a), arise from long-wavelength extrinsic strain[1,2]. The interlayer displacement at the domain wall network nodes can be calculated from the deviation of the node position from the unstrained lattice, as indicated with arrows in (a). From this interlayer displacement **u(r)**, the strain tensor

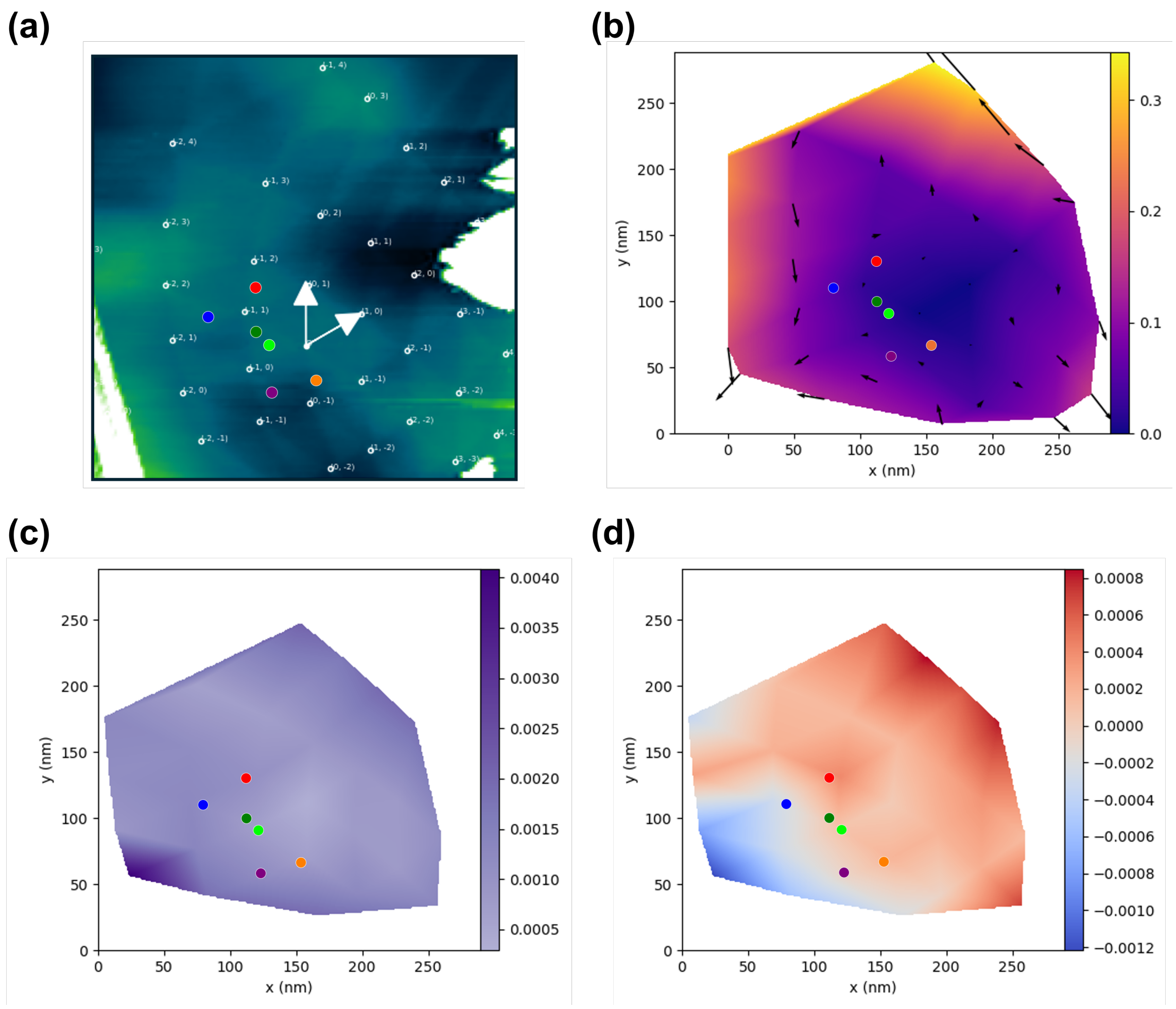

Supplementary Figure 1: (a) STM topographic map ($V_b$=250mV; $I_t$=350pA), with vertices label with their indices. Unstrained triangle indicted with arrows. (b) Interlayer displacement magnitude and direction. (c) Hydrostatic strain. (d) Shear strain.

components are computed via spatial derivatives of the displacement, $\varepsilon_{ij} = (\partial_i u_j + \partial_j u_i)/2$. Figure (b) shows the magnitude and direction of the interlayer displacement, providing a direct visualization of the strain-induced lattice distortion. The trace and off-diagonal components of the strain tensor correspond to the hydrostatic and shear strain, respectively, representing local isotropic and anisotropic lattice deformations, as shown in (c) and (d)

**Supplementary Note 2: Apparent height in three stacking regions**

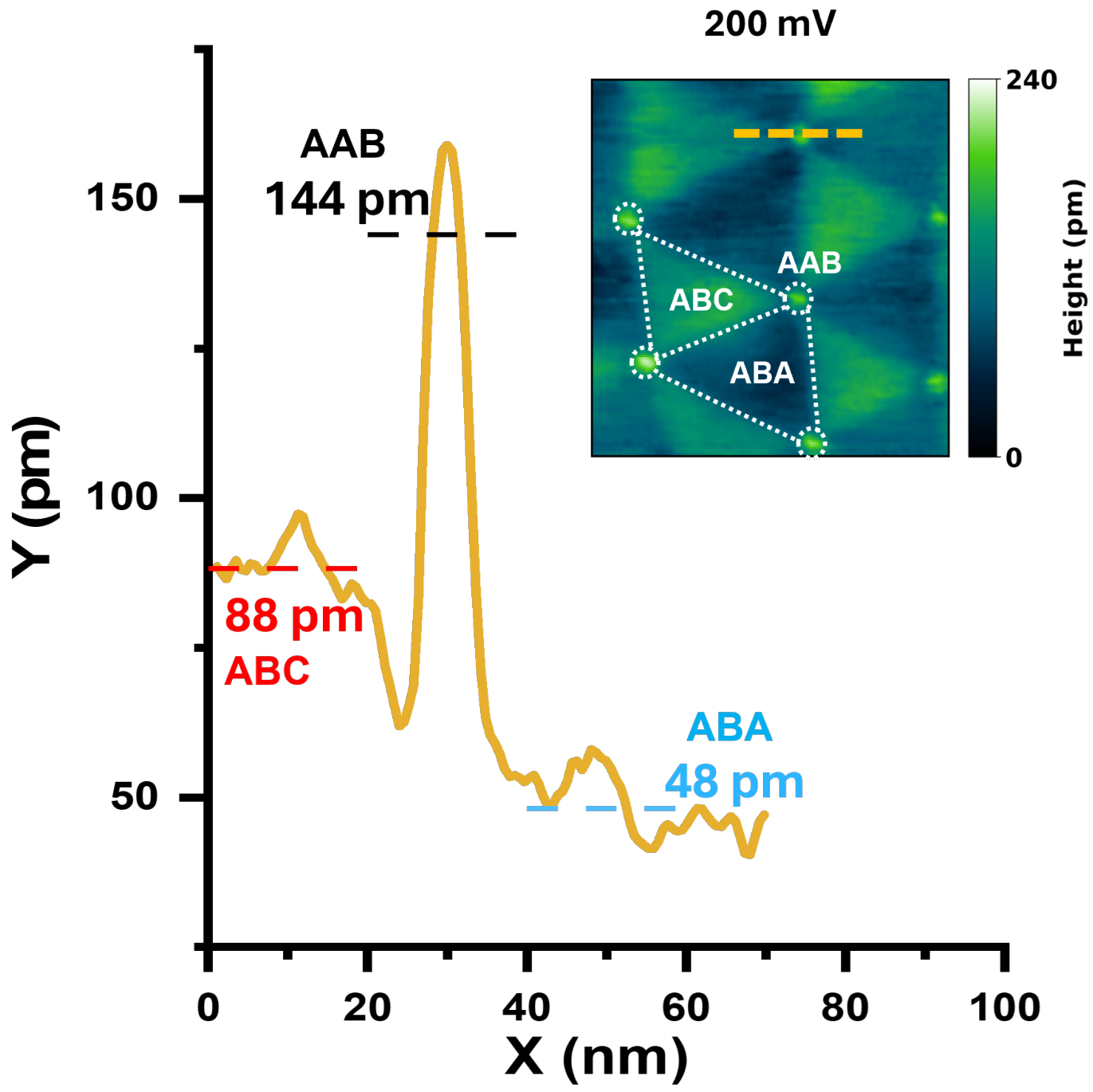

Supplementary Figure 2: STM topography height line profile corresponding to the line in the 200mV map in Figure 4. The inset shows the location on the sample.

It should be noted that the STM height profile represents an apparent height, arising from a convolution of the physical surface corrugation and variations in the LDOS at the applied sample bias.

## Supplementary Note 3: Band structure calculations

Band structure calculations were performed using the hybrid $k \cdot p$ tight-binding model for multilayer graphene[3]. The Hamiltonian for trilayer graphene is constructed by considering the interface between layers 1 and 2 as Bernal stacked, and layers 2 and 3 have a twisted interface, parameterized in terms of the lateral offset between nearest A sublattices, $\boldsymbol{r}_0$. In the basis of A and B sublattices, $\{\psi^{A1}, \psi^{B1}, \psi^{A2}, \psi^{B2}, \psi^{A3}, \psi^{B3}\}$, the Hamiltonian reads as

$$H_{\mathrm{kpTB}} = \begin{pmatrix} H_1^g & V^\dagger & W(\boldsymbol{r_0}) \\ V & H_2^g & V_{\mathrm{int}}(\boldsymbol{r_0}) \\ W^\dagger(\boldsymbol{r_0}) & V_{\mathrm{int}}^\dagger(\boldsymbol{r}_0) & H_3^g \end{pmatrix},$$

with

$$H_1^g = H_g + \begin{pmatrix} \Delta_{hBN} & 0 \\ 0 & \Delta' + \Delta_{hBN} \end{pmatrix}, \qquad H_2^g = H_g + \begin{pmatrix} \Delta' & 0 \\ 0 & 0 \end{pmatrix},$$

$$H_3^g = H_g + \begin{pmatrix} \Delta_{vac} & 0 \\ 0 & \Delta_{vac} \end{pmatrix},$$

$$H_g = v \begin{pmatrix} 0 & \pi_\xi^\dagger \\ \pi_\xi & 0 \end{pmatrix}, \qquad \pi_\xi = \xi p_x + i p_y,$$

$$V = \begin{pmatrix} -v_4 \pi_\xi & \gamma_1 \\ -v_3 \pi_\xi^\dagger & -v_4 \pi_\xi \end{pmatrix},$$

$$V_{\mathrm{int}}(\boldsymbol{r_0}) = \sum_{j=0}^{2} \Bigg[ \left[ \frac{\gamma_1}{3} - \frac{2 v_4 \hbar}{3K} \boldsymbol{k} \cdot \boldsymbol{K}_\xi^{(j)} \right] \begin{pmatrix} 1 & e^{i\xi \frac{2\pi}{3} j} \\ e^{-i\xi\frac{2\pi}{3} j} & 1 \end{pmatrix} + \frac{\xi\, 2(v_3 - v_4)\hbar}{3K} \left( \boldsymbol{k} \times \boldsymbol{K}_\xi^{(j)} \right)_z \begin{pmatrix} 0 & i e^{i\xi\frac{2\pi}{3} j} \\ -i e^{-i\xi \frac{2\pi}{3} j} & 0 \end{pmatrix} \Bigg] e^{-i \boldsymbol{K}_\xi^{(j)} \cdot \boldsymbol{r_0}},$$

$$W(\boldsymbol{r_0}) = \frac{1}{6} \sum_{j=0}^{2} \begin{pmatrix} \gamma_2 e^{+i\xi\frac{2\pi}{3} j} & \gamma_2 e^{-i\xi \frac{2\pi}{3} j} \\ \gamma_5 & \gamma_5 e^{+i\xi\frac{2\pi}{3} j} \end{pmatrix} e^{-i \boldsymbol{K}_\xi^{(j)} \cdot \boldsymbol{r}_0}.$$

Here, we use the Sloczewski-Weiss-McClure (SWMcC) parameters $v \approx 1.02 \cdot 10^6 m/s$, $\gamma_1 =$ 390meV, $\gamma_2 = 17$meV, $\gamma_5 = 38$meV, $v_3 \approx 0.1v$, $v_4 \approx 0.022v$, $\Delta' = 25$meV, and $\xi = \pm 1$ is the valley index. The wave vectors of the $K$ valleys are $\boldsymbol{K}_\xi^{(j)} = \xi \frac{4\pi}{3a} \left[\cos\left(\frac{2\pi j}{3}\right), -\sin\left(\frac{2\pi j}{3}\right)\right]$, with $a =$ 0.246nm. For $\boldsymbol{r}_{\mathbf{0}}^{(\mathrm{ABC})} = \frac{a}{\sqrt{3}}\hat{\boldsymbol{y}}$, $\boldsymbol{r}_{\mathbf{0}}^{(\mathrm{ABA})} = -\frac{a}{\sqrt{3}}\hat{\boldsymbol{y}}$ and $\boldsymbol{r}_{\mathbf{0}}^{(\mathrm{AAB})} = \mathbf{0}$ the Hamiltonian above corresponds to ABC, ABA and AAB stacking, respectively.

**Supplementary Note 4: Thermal broadening at room temperature**

The energy resolution in STS at finite temperature is limited by the thermal broadening of the Fermi–Dirac distribution[4]. The measured dI/dV is given by a convolution of the sample density of states with the derivative of the Fermi–Dirac distribution. As a result, thermal effects introduce an intrinsic energy broadening that can be approximated by $\Delta \mathrm{E}_{\mathrm{thermal}} \approx 3.2 k_B T$, where $k_B$ is the Boltzmann constant and T is the temperature.

For room temperature measurements at T = 298 K, the thermal broadening is calculated as follows: $\Delta \mathrm{E}_{\mathrm{thermal}} \approx 3.2 k_B T = 3.2 \times (8.617 \times 10^{-5}\ eV/K) \times\ 298K = 82.2\ meV$.

**Supplementary Note 5: Strain-induced pseudomagnetic field**

The strain textures arising from lattice reconstruction effects translate into a vector gauge field in the graphene Hamiltonian. This vector potential in valleys $\xi = \pm 1$ is[5]

$$\boldsymbol{A}_{\xi}^{(\ell)}(\boldsymbol{r}) = \xi \frac{\sqrt{3}\hbar\eta_0}{2ea} \begin{pmatrix} u_{xx}^{(\ell)}(\boldsymbol{r}) - u_{yy}^{(\ell)}(\boldsymbol{r}) \\ -2u_{xy}^{(\ell)}(\boldsymbol{r}) \end{pmatrix}, \tag{1}$$

where $\eta_0 \approx -3$ is the change of the nearest neighbour coupling between carbon atoms in graphene due to the change of the bond length[6]. The corresponding pseudomagnetic field (PMF) is

$$\begin{aligned} B_{\xi}^{(\ell)}(\boldsymbol{r}) &= \left[\nabla \times \boldsymbol{A}_{\xi}^{(\ell)}(\boldsymbol{r})\right]_z \\ &= -\xi \frac{\sqrt{3}\hbar\eta_0}{2ea}\left[2\partial_x u_{xy}^{(\ell)}(\boldsymbol{r}) + \partial_y \left(u_{xx}^{(\ell)}(\boldsymbol{r}) - u_{yy}^{(\ell)}(\boldsymbol{r})\right)\right]. \end{aligned} \tag{2}$$

Note that the maximum value of PMF in layer 3 (~20T) corresponds to a magnetic length of $L \approx$ 6nm. However, the domain wall width of ~7nm estimated in the Main text indicates that the formation of pseudo-Landau levels can be ruled out since any possible localization spot has a vanishingly small average PMF around a disk of radius $L$.

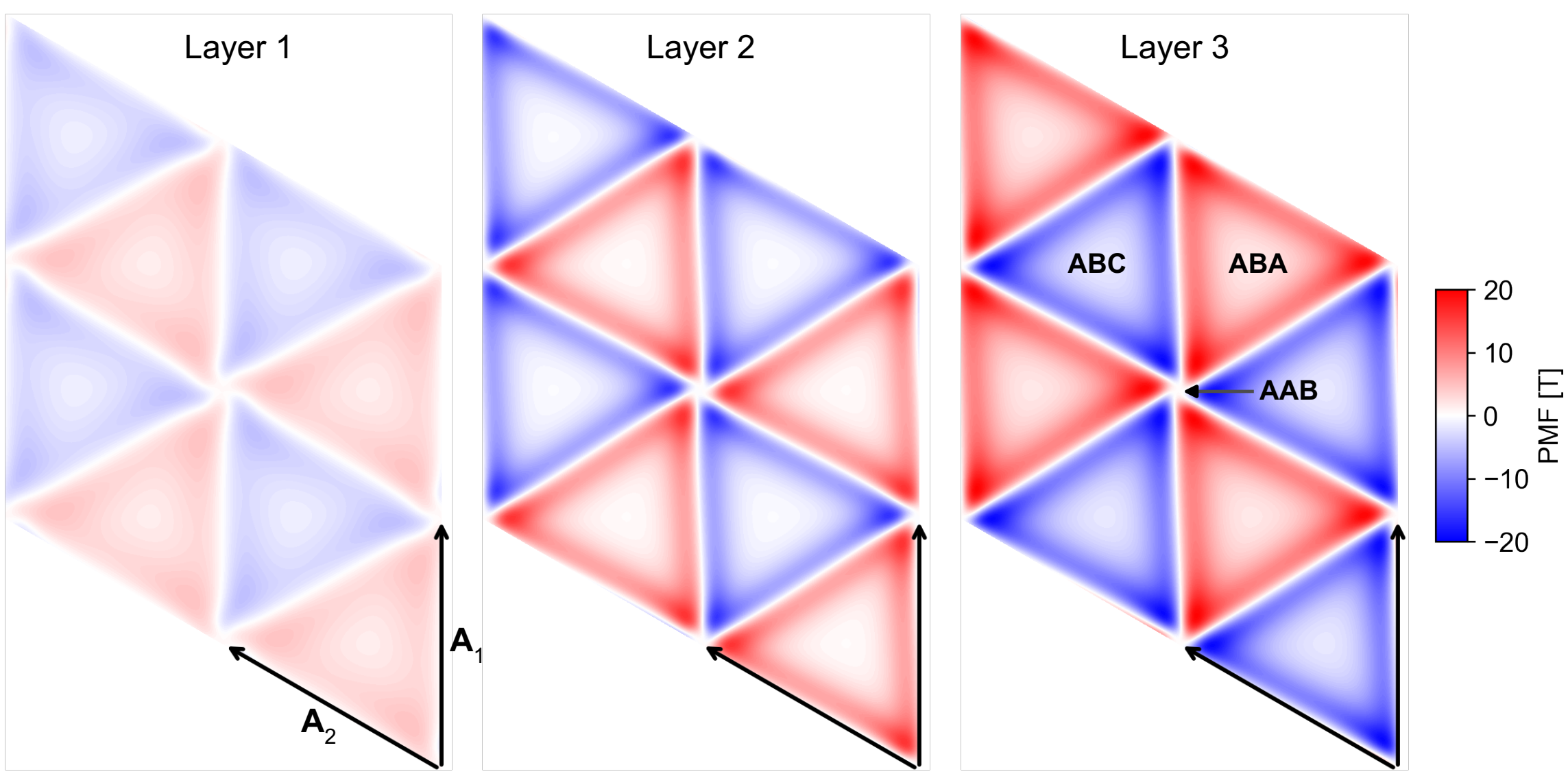

Supplementary Figure 3: Strain-induced pseudomagnetic field induced in each layer of trilayer graphene with the topmost layer twisted by an angle $\theta = 0.3°$.

**Supplementary Note 6: First-principles density of states (DOS) calculations**

First-principles calculations of the DOS, corresponding to the theoretical band structures presented in Figure 2(b) of the main text, were performed for the three stacking domains using a twist angle of 0.3° to match the experimental sample. The simulation was for 0 K without thermal broadening.

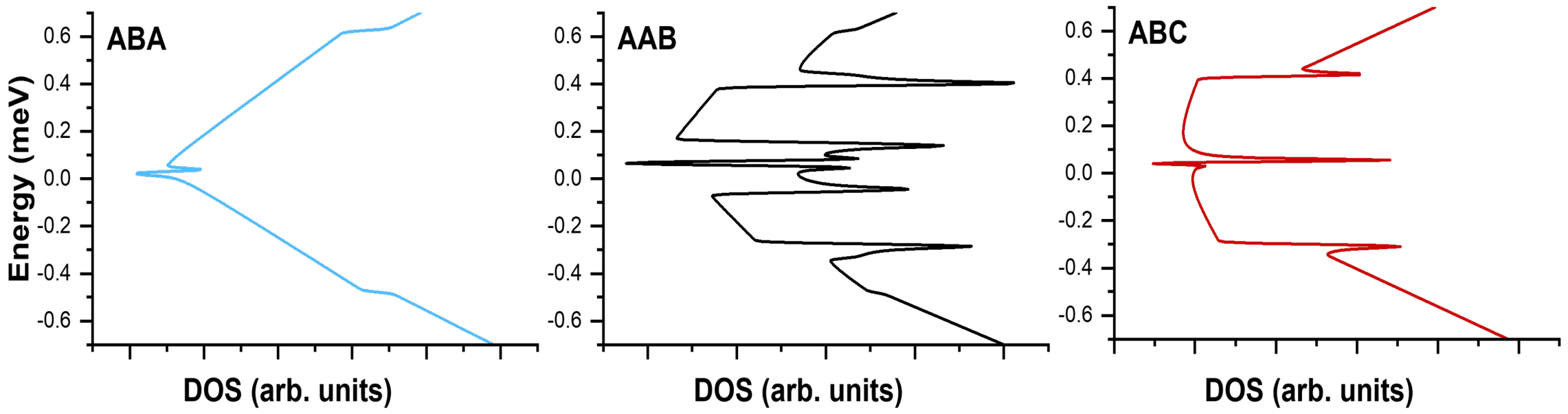

Supplementary Figure 4: First principles density of states of three stacking domains.

**Supplementary Note 7: Displacement fields and strain maps from mesoscale lattice relaxation model**

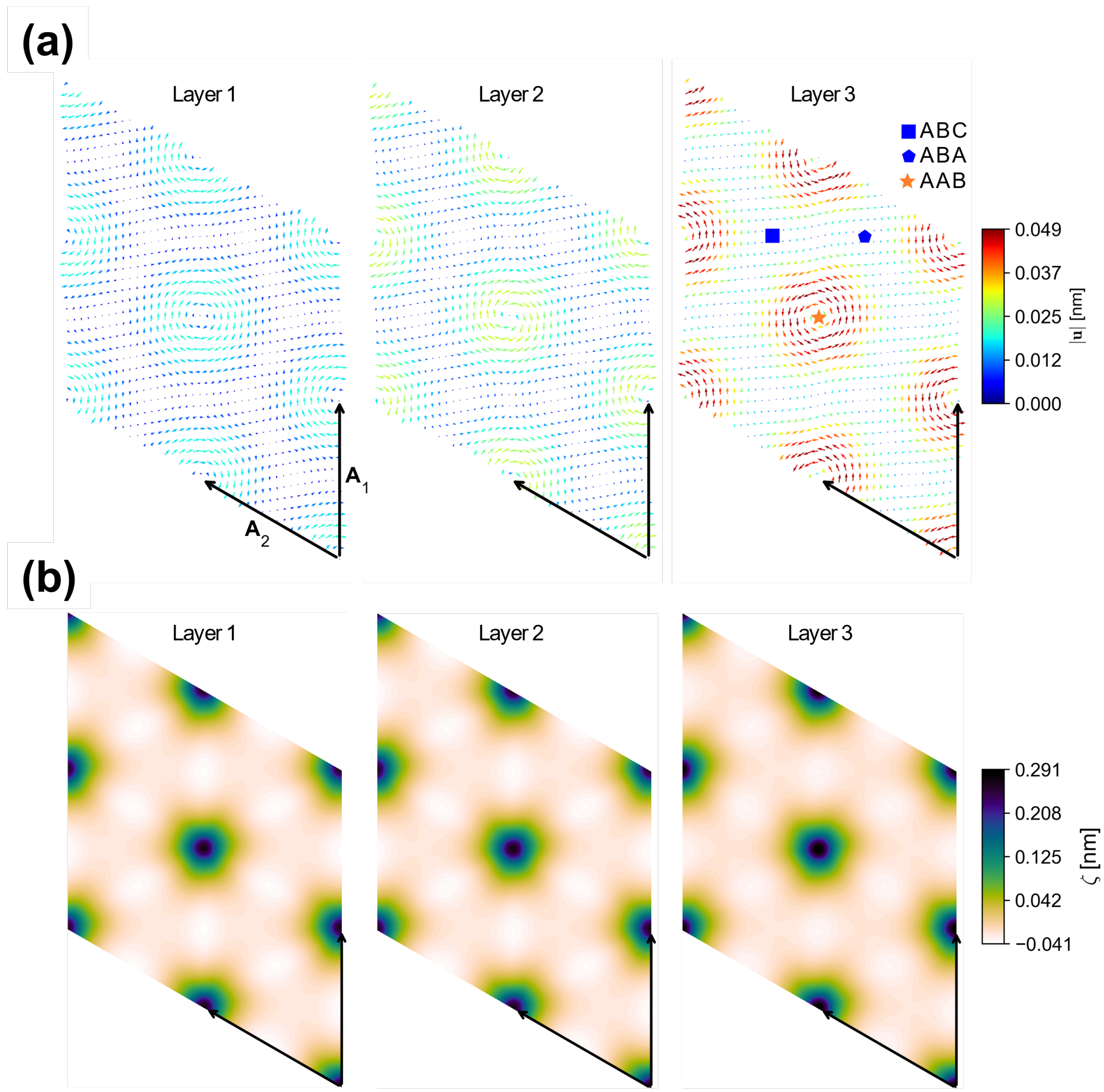

Supplementary Figure 5: Displacement fields (a) $\mathbf{u}^{(\ell)}$ and (b) $\zeta^{(\ell)}$ resulting from lattice relaxation in trilayer graphene with the topmost layer twisted by an angle $\theta = 0.3°$. The moiré superlattice vectors are shown in the first panel, where $|\mathbf{A}_{1,2}| \approx a/\theta = 46.98\text{nm}$. Symbols in the upper panel indicate the high symmetry stacking of the twisted trilayer.

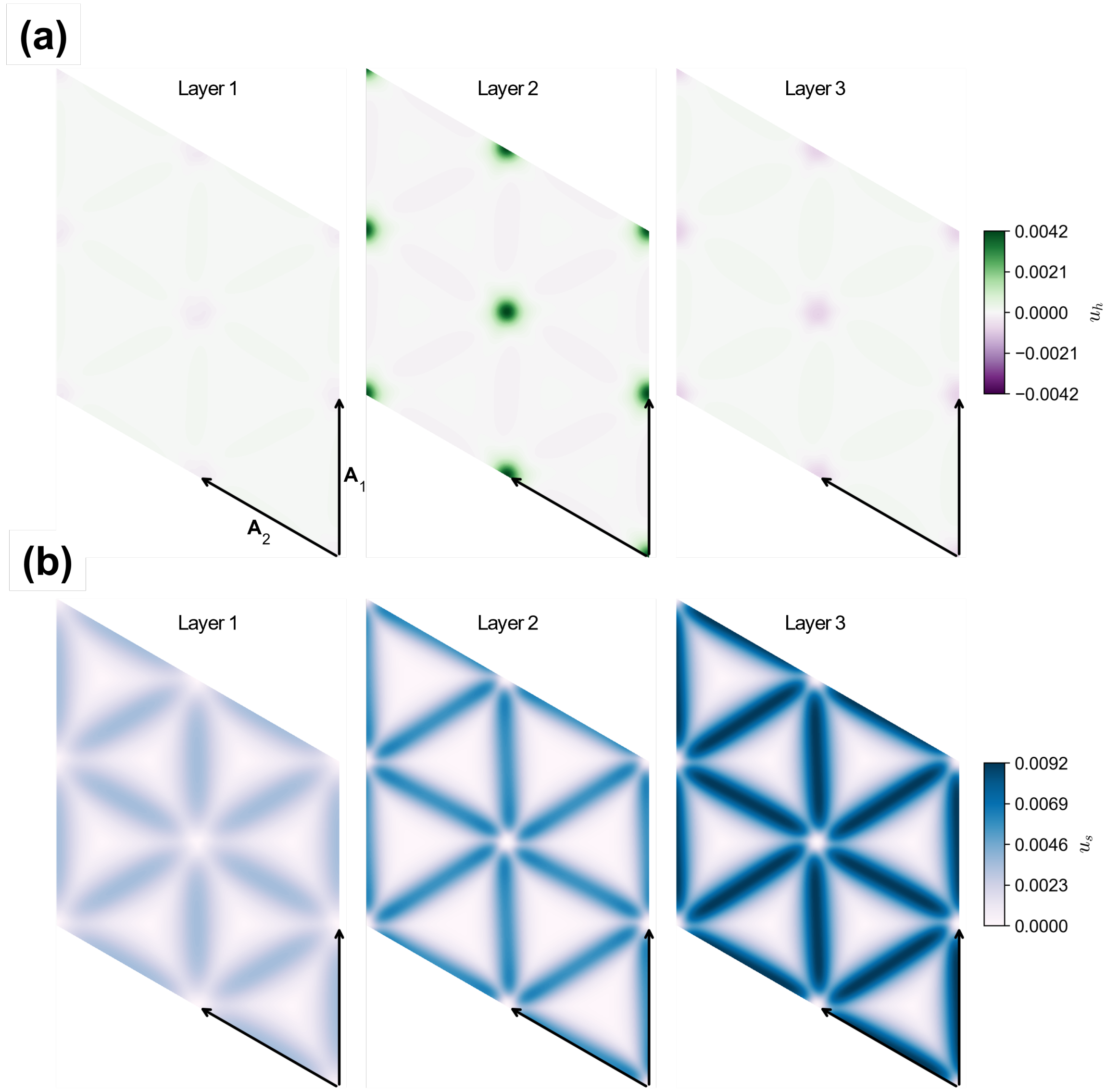

Supplementary Figure 6: (a) Hydrostatic, $\mathrm{u_h} = \mathrm{Tr}(\mathrm{u_{ij}})$, and (b) shear, $\mathrm{u_s} = \sqrt{(\mathrm{u_{xx}} - \mathrm{u_{yy}})^2 + 4\mathrm{u_{xy}^2}}$, components of strain in trilayer graphene with the topmost layer twisted by an angle $\theta = 0.3°$.

## Supplementary Note 8: Extraction of domain wall width from $dI/dV$ maps

We define the width of the domain walls in the dI/dV maps as the distance between the regions of constant dI/dV. To determine this distance, we segment each dI/dV map into triangles. This segmentation is accomplished via the following processing pipeline,

① Take the gradient of a smoothed dI/dV map.

② Threshold the resulting image to include only the regions where the dI/dV is not constant

③ Skeletonize the result to produce an approximation of the domain wall network and label each bounded region, creating masks of each triangle.

For each mask, a distance map, as shown in (b), is generated that represents the distance of each pixel within each triangle to the nearest domain wall. Plotting the dI/dV values measured at each pixel as a function of the distance to the nearest domain wall yields a curve in (c) that resembles an erf function. We define the domain wall width as the full-width-half-max (FWHM) of the derivative of the erf function fit to this curve.

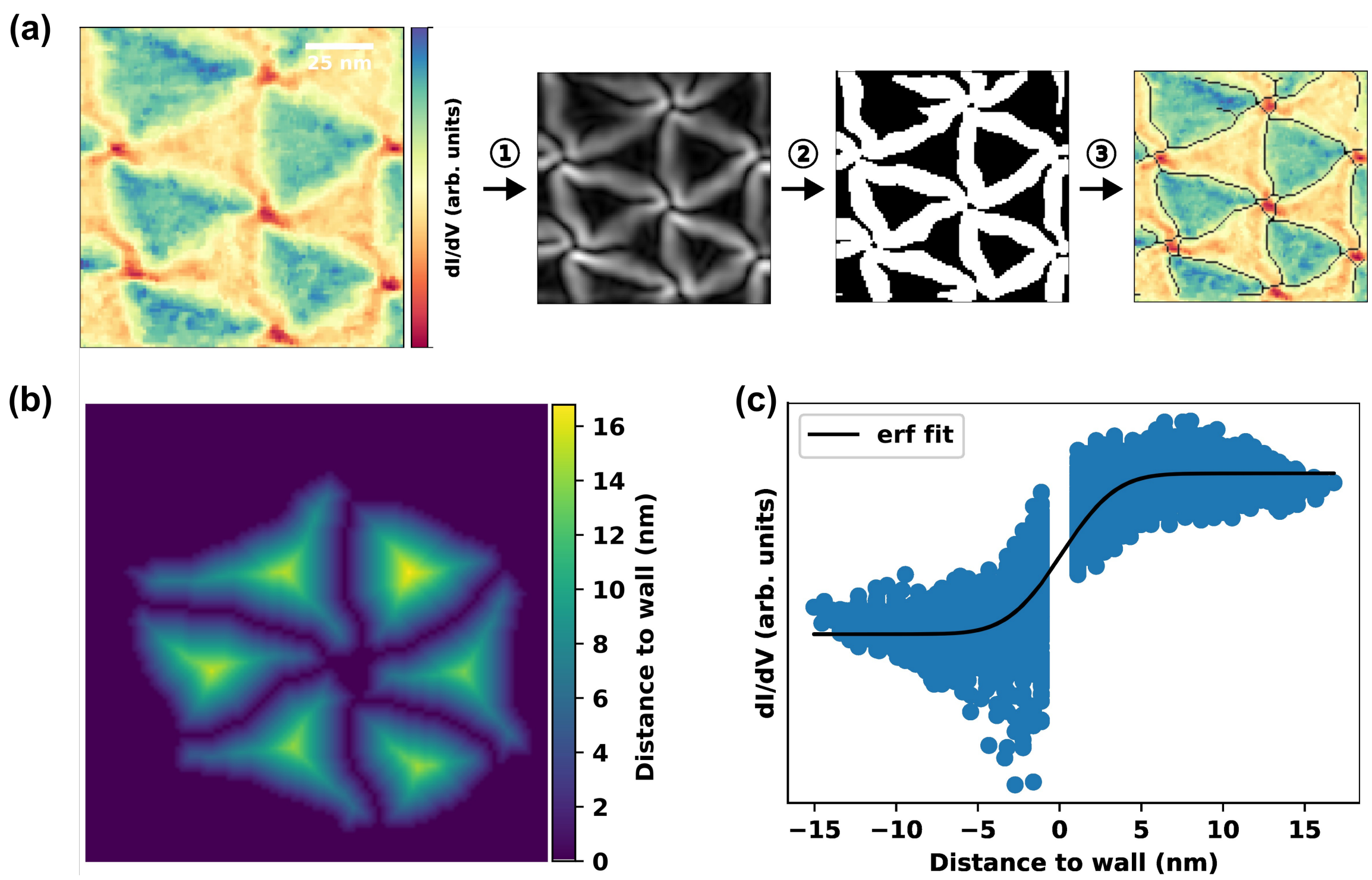

Supplementary Figure 7: (a) Processing pipeline for labelling each triangle in dI/dV map. ① Starting from the raw data, the gradient of the image is calculated. ② Threshold to isolate the domain walls. ③ Skeletonized to produce an approximation of the domain wall network. (b) A distance map representing the distance from each pixel within a triangle to the nearest domain wall. (c) The dI/dV value for each pixel as a function of the distance of each pixel to the nearest domain wall.

**Supplementary Note 9: Additional STS measurements on multiple regions of sample**

Supplementary STS measurements were acquired on additional regions of the same sample in three stacking domains. All spectra exhibit the same characteristic feature as those shown in Figure 2(a) of the main text, demonstrating that the observed features are spatially uniform and robust across different domains.

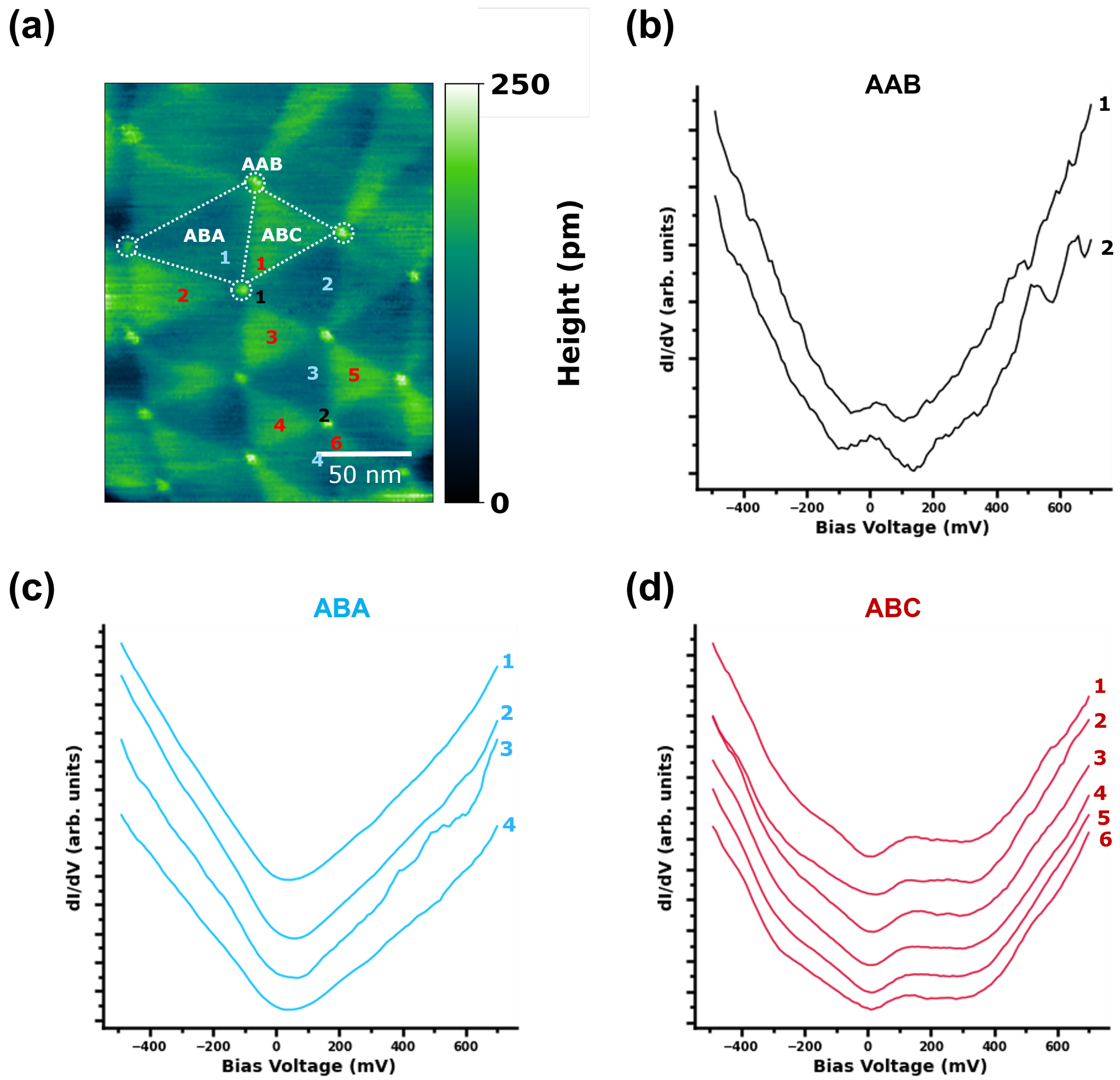

Supplementary Figure 8: (a) STM topograph showing triangular reconstruction domains (VB = 250mV; It = 350pA). The different stackings identified are indicated within the dashed lines. (b)(c)(d) Scanning tunnelling spectra (STS) acquired on spectral locations labelled in (a). Individual spectra in the figures are vertically offset for clarity. Spectra tunnelling junction settings: R=2.00GΩ, It=150pA (ABA, ABC), R=1.67GΩ; It=150pA (ABA).

**Supplementary Note 10: Mesoscale lattice relaxation calculation**

The modeling of atomic relaxation in trilayer graphene with the topmost layer twisted by an angle $\theta$ is performed through a multiscale approach [*Physical review letters*, *124*(20), 206101 (2020); *Nature nanotechnology*, *15*(7), 592-597 (2020)] based on an interpolated adhesion energy density, dependent on both the vertical stacking and interlayer distance [*Phys. Rev. B*, *92*(15), 155438 (2015)], and continuum elasticity theory. Within this approach, we introduce the in-plane, $\boldsymbol{u}^{(\ell)}$, and out-of-plane displacement fields, $\zeta^{(\ell)}$, in each layer ($\ell = 1, 2, 3$), such that these minimize the total energy of the moiré unit cell (mUC),

$$\mathcal{H} = \int_{\mathrm{mUC}} \mathrm{d}^2 r \left[ W\left(\boldsymbol{r}_0^{(1,2)}, d^{(1,2)}\right) + W\left(\boldsymbol{r}_0^{(2,3)}, d^{(2,3)}\right) + \sum_{\ell=1}^{3} \{U_\ell + B_\ell\} \right]. \quad (1)$$

where

$$U_\ell = \frac{\lambda}{2}\left(u_{ii}^{(\ell)}(\boldsymbol{r})\right)^2 + \mu u_{ij}^{(\ell)}(\boldsymbol{r}) u_{ji}^{(\ell)}(\boldsymbol{r}),$$

$$B_\ell = \frac{\kappa}{2}\left[\left(\nabla^2 \zeta^{(\ell)}(\boldsymbol{r})\right)^2 + 2(1-\sigma)\left\{\left(\partial_x \partial_y \zeta^{(\ell)}(\boldsymbol{r})\right)^2 - \left(\partial_x^2 \zeta^{(\ell)}(\boldsymbol{r})\right)\left(\partial_y^2 \zeta^{(\ell)}(\boldsymbol{r})\right)\right\}\right]. \quad (2)$$

In Eq. (1), $W$ is the adhesion energy density, accounted for both interfaces between graphene monolayers, and is determined by the lateral offset between A sublattices in neighboring layers, $\boldsymbol{r}_0$, and the interlayer distance $d$. Explicit functional form of $W(\boldsymbol{r}_0, d)$, parametrized from density functional theory (DFT) calculations, can be found in Ref. [*Phys. Rev. B*, *92*(15), 155438 (2015)]. The terms $U_\ell$ and $B_\ell$ are the elastic and bending energy densities, given in terms of the strain tensor $u_{ij}^{(\ell)}(\boldsymbol{r}) = \frac{1}{2}\left[\partial_i u_j^{(\ell)}(\boldsymbol{r}) + \partial_j u_i^{(\ell)}(\boldsymbol{r}) + \left(\partial_i \zeta^{(\ell)}(\boldsymbol{r})\right)\left(\partial_j \zeta^{(\ell)}(\boldsymbol{r})\right)\right]$, first Lamé coefficient $\lambda = 450\mathrm{eV/nm}^2$, shear moduli $\mu = 920\mathrm{eV/nm}^2$, bending rigidity $\kappa = 1.44\mathrm{eV}$, and Poisson ratio $\sigma = 0.165$ [*Journal of applied physics*, *41*(8), 3373-3382 (1970); *Nano letters, 13(1), 26-30* (2013)]. The lateral offset vectors for the AB stacked ($\ell = 1$ and 2) and the twisted interface ($\ell = 2$ and 3) are

$$\boldsymbol{r}_0^{(1,2)}(\boldsymbol{r}) = \boldsymbol{r}_0^{(\mathrm{AB})} + \boldsymbol{u}^{(2)}(\boldsymbol{r}) - \boldsymbol{u}^{(1)}(\boldsymbol{r}) + \frac{d^{(1,2)}(\boldsymbol{r})}{2} \nabla\left[\zeta^{(1)}(\boldsymbol{r}) + \zeta^{(2)}(\boldsymbol{r})\right], \quad (3)$$

$$\boldsymbol{r}_0^{(2,3)}(\boldsymbol{r}) = \theta\hat{\boldsymbol{z}} \times \boldsymbol{r} + \boldsymbol{u}^{(3)}(\boldsymbol{r}) - \boldsymbol{u}^{(2)}(\boldsymbol{r}) + \frac{d^{(2,3)}(\boldsymbol{r})}{2}\nabla\left[\zeta^{(2)}(\boldsymbol{r}) + \zeta^{(3)}(\boldsymbol{r})\right],$$

where $\boldsymbol{r}_0^{(\mathrm{AB})} = \frac{a}{\sqrt{3}}\hat{\boldsymbol{y}}$. Similarly, the interlayer distances are

$$d^{(k,l)}(\boldsymbol{r}) = d_0 - \zeta^{(k)}(\boldsymbol{r}) + \zeta^{(l)}(\boldsymbol{r}), \tag{4}$$

with $d_0 = 0.339\text{nm}$ an interlayer distance reference. The fourth term in Eqs. (3) is an additional in-plane offset shift due to bending of the layers, which plays an important role for non-encapsulated structures (see Supplementary Figure ***). Note that bending the layers in opposite directions cancels the contribution of this term, while bending in the same direction results in an effective lateral offset, contributing to atomic reconstruction effects. Furthermore, given the low bending rigidity of graphene monolayers, vertical displacement of the layers in the same direction results in a drastic reduction of the AA stacking node area, accompanied by twirling of domain walls around this region.

The optimal configuration of the displacement fields was found by defining these on a grid on the mUC, where the spacing between points was reduced until achieving convergence. The minimal energy was determined using the limited-memory Broyden-Fletcher-Goldfarb-Shanno algorithm (L-BFGS) with a tolerance of $10^{-11}$eV. The results for displacements fields in the trilayer system with $\theta = 0.3°$ are shown in Supplementary Figure ***.

**Supplementary Note 11: In-plane offset correction due to layer tilting**

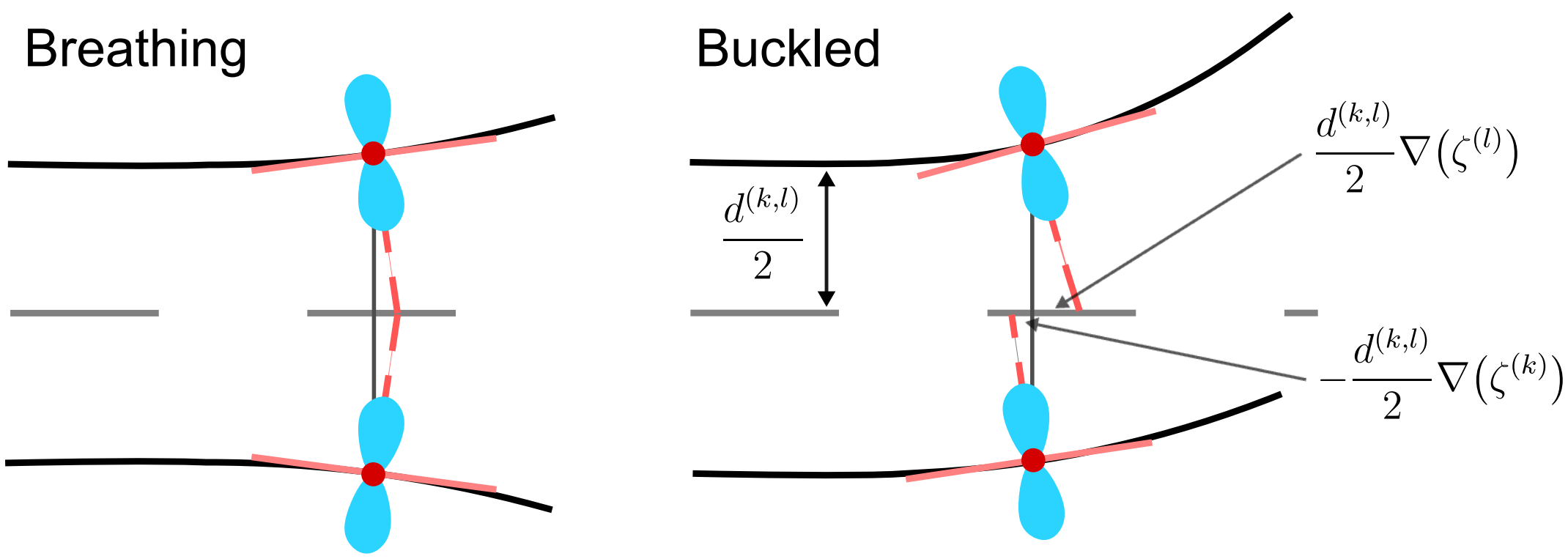

Supplementary Figure 9: Effective correction to the lateral offset from bending of the layers. In the breathing mode (left panel), the lateral offset correction is canceled due to bending in opposite directions. The buckled mode (right panel) shows an effective in-plane offset result of the bending of the layers in the same direction.